\input harvmac.tex

\lref\malda{J.~ M.~ Maldacena, "The Large N Limit of Superconformal
Field Theories and Supergravity", hep-th/9711200.}%
\lref\witholo{E.~Witten, "Anti De Sitter Space And Holography",
hep-th/9802150.}
\lref\witthe{E.~Witten, ``Anti-de Sitter Space, Thermal Phase Transition,
And Confinement In Gauge Theories'', hep-th/9803131.}
\lref\w{E.~Witten,"Baryons And Branes In Anti de Sitter Space",
hep-th/9805112.}%
\lref\ks{S.~Kachru, E.~ Silverstein, "4d Conformal Field Theories
and Strings on Orbifolds", hep-th/9802183.}%
\lref\lnv{A.~ Lawrence, N.~ Nekrasov, C.~ Vafa,
"On Conformal Theories in Four Dimensions", hep-th/9803015.}%
\lref\bkv{M.~Bershadsky, Z.~Kakushadze, C.~Vafa, "String Expansion
as Large N Expansion of Gauge Theories", hep-th/9803076.}
\lref\dgmr{M.~R.~Douglas, G.~Moore, "D-branes, Quivers, and ALE
Instantons", hep-th/9603167.}
\lref\dhvw{L.~Dixon, J.~A.~Harvey, C.~Vafa and E.~Witten,
\np B 261 (1985) 678; \np B 274 (1986) 285.}
\lref\dgm{M.~R.~ Douglas, B.~ R.~ Greene, D.~ R.~ Morrison, "Orbifold
Resolution by D-Branes", Nucl.Phys. {\bf B506} (1997) 84.}
\lref\bj{M.~Bershadsky, A.~Johansen "Large N limit of orbifold
field theories", hep-th/9803249.}
\lref\gv{R.~ Gopakumar, C.~Vafa, "Branes and Fundamental Groups",
hep-th/9712048.}%
\lref\witbound{E.~Witten, ``Bound States Of Strings And $p$-Branes'',
hep-th/9510135.}
\lref\go{D.~J.~ Gross, H.~ Ooguri, "Aspects of Large N Gauge
Theory Dynamics as Seen by String Theory", hep-th/9805129.}%
\lref\toz{Y.~ Oz, J.~ Terning, "Orbifolds of $AdS_5 \times S^5$
and 4d Conformal Field Theories", hep-th/9803167.}%
\lref\hu{A.~ Hanany, A.~ M.~ Uranga, "Brane Boxes and Branes
on Singularities", hep-th/9805139.}
\lref\gkp{S.~S.~ Gubser, I.~R.~ Klebanov, A.~M.~ Polyakov,
"Gauge Theory Correlators from Non-Critical String Theory",
hep-th/9802109.}
\lref\gun{M.~Gunyadin, L.~J.~Romans, N.~P.~Warner,
Nucl.Phys. {\bf B272} (1986) 598; Phys. Lett. {\bf B154} (1985) 268.}
\lref\shiraz{S.~ Minwalla, "Restrictions Imposed by Superconformal
Invariance on Quantum Field Theories", hep-th/9712074.}
\lref\aoy{O.~ Aharony, Y.~ Oz, Z.~ Yin, "M Theory on $AdS_p \times S^{11-p}$
and Superconformal Field Theories", hep-th/9803051.}
\lref\ffz{S.~ Ferrara, C.~ Fronsdal, A.~ Zaffaroni,
"On N=8 Supergravity on $AdS_5$ and N=4 Superconformal Yang-Mills
theory", hep-th/9802203.}
\lref\fz{S. ~Ferrara, A.~ Zaffaroni, "N=1,2 4D Superconformal Field
Theories and Supergravity in $AdS_5$", hep-th/9803060.}
\lref\jm{C.~ V.~ Johnson, R.~ C.~ Myers, Phys.Rev. {\bf D55} (1997) 6382.}
\lref\sen{A.~Sen, "Dynamics of Multiple Kaluza-Klein Monopoles
in M- and String Theory", Adv.Theor.Math.Phys. {\bf 1} (1998) 115;
"A Note on Enhanced Gauge Symmetries in M- and String Theory",
hep-th/9707123.}
\lref\gh{G.~W.~Gibbons, S.~W.~Hawking, Commun. Math. Phys.
{\bf 66} (1979) 291.}
\lref\ruback{P.~J.~Ruback, Commun. Math. Phys. {\bf 107} (1986) 93.}
\lref\ghlam{G.~W.~Gibbons, S.~W.~Hawking, Phys. Rev. {\bf D15}
(1977) 2738.}
\lref\monopol{N.~S.~Manton, Phys.Lett. {\bf B110} (1982) 54.}
\lref\mon{M.~F.~Atiyah, N.~J.~Hitchin, Phys.Lett. {\bf A107} (1985) 21.}
\lref\afm{O.~ Aharony, A.~ Fayyazuddin, J.~ Maldacena, "The Large N
Limit of ${\cal N}=2,1$ Field Theories from Threebranes in
F-theory", hep-th/9806159.}
\lref\dima{G.~ Dall'Agata, K.~ Lechner, D.~ Sorokin,
Class.Quant.Grav. {\bf 14} (1997) L195.}

\lref\crocodile{A.~ Fomenko, D.~ Fuks, ``The course in homotopy topology''}
\def\IB{\relax\hbox{$\inbar\kern-.3em{\rm B}$}}
\def\IC{\relax\hbox{$\inbar\kern-.3em{\rm C}$}}
\def\ID{\relax\hbox{$\inbar\kern-.3em{\rm D}$}}
\def\IE{\relax\hbox{$\inbar\kern-.3em{\rm E}$}}
\def\IF{\relax\hbox{$\inbar\kern-.3em{\rm F}$}}
\def\IG{\relax\hbox{$\inbar\kern-.3em{\rm G}$}}
\def\IGa{\relax\hbox{${\rm I}\kern-.18em\Gamma$}}
\def\IH{\relax{\rm I\kern-.18em H}}
\def\IK{\relax{\rm I\kern-.18em K}}
\def\IL{\relax{\rm I\kern-.18em L}}
\def\IP{\relax{\rm I\kern-.18em P}}
\def\IR{\relax{\rm I\kern-.18em R}}
\def\IZ{\relax\ifmmode\mathchoice
{\hbox{\cmss Z\kern-.4em Z}}{\hbox{\cmss Z\kern-.4em Z}}
{\lower.9pt\hbox{\cmsss Z\kern-.4em Z}}
{\lower1.2pt\hbox{\cmsss Z\kern-.4em Z}}\else{\cmss Z\kern-.4em
Z}\fi}

\def\II{\relax{\rm I\kern-.18em I}}


\def\CB {{\cal B}}

\def\CL {{\cal L}}
\def\CM {{\cal M}}
\def\CN {{\cal N}}
\def\CO {{\cal O}}


\def\p{\partial}



\def\Tr{{\rm Tr}}

\def\p{\partial}

\def\inbar{\,\vrule height1.5ex width.4pt depth0pt}
\font\cmss=cmss10 \font\cmsss=cmss10 at 7pt

\def\a{\alpha}

\def\b{\beta}
\def\g{\gamma}
\def\d{\delta}

\def\la{\lambda}
\def\th{\theta}

\def\om{\omega}

\def\p{\partial}
\def\R{\relax{\rm I\kern-.18em R}}
\font\cmss=cmss10 \font\cmsss=cmss10 at 7pt
\def\Z{\relax\ifmmode\mathchoice
{\hbox{\cmss Z\kern-.4em Z}}{\hbox{\cmss Z\kern-.4em Z}}
{\lower.9pt\hbox{\cmsss Z\kern-.4em Z}}
{\lower1.2pt\hbox{\cmsss Z\kern-.4em Z}}\else{\cmss Z\kern-.4em Z}\fi}

\def\np{{Nucl. Phys. }}

\def\CM{{\cal M}}

\def\vp{\varphi}

\hbox{PUPT-1800}
\hbox{ITEP-TH-33/98}
\hbox{Landau-98-TMP-2}
  \Title{ \vbox{\baselineskip12pt\hbox{hep-th/9806180}}}
{\vbox{ \centerline
{Comments on $\CN=2$ $AdS$ Orbifolds}
\vskip2pt
    }}
\centerline{Sergei Gukov\foot{On leave from the Institute of
Theoretical and Experimental Physics and the L.D.~Landau
Institute for Theoretical Physics}}
\vskip 2pt
\centerline{Joseph Henry Laboratories, Princeton University}
\centerline{Princeton, New Jersey 08544}
\centerline{gukov@pupgg.princeton.edu}
\vskip 30pt

{\bf \centerline{Abstract}}

We discuss twisted states of AdS orbifolds
which couple to $\CN=2$ chiral primary operators
{\it not} invariant under exchange of the gauge factors.
Kaluza-Klein reduction on the fixed circle
gives the correct conformal dimensions
of operators in the superconformal theory and
involves some aspects of monopole dynamics
in the non-trivial background.
As a byproduct we found evidence for decoupling of
$U(1)$ factors in the four-dimensional gauge theory.

\Date{June 1998}
\newsec{Introduction}

Since the remarkable conjecture of Maldacena \malda\ a lot of progress
has been made towards understanding the dynamics of superconformal
field theories (SCFT) in four dimensions. According to \malda,
$\CN=4$ SCFT is dual to type IIB string theory compactified
on anti-de Sitter ($AdS$) space of the form $AdS_5 \times S^5$.
To make this relation more precise one has to compare the states
in both theories and their interactions \refs{\witholo,\gkp}.
Correlation functions of the fields in the
four-dimensional theory on the boundary can be evaluated via
the asymptotic dependence of the supergravity action.

The conjecture was extended to the theories with lower number of
supersymmetries by means of the  orbifold construction
\refs{\ks,\lnv}.
The idea is to place $N$ three-branes at the orbifold point
of $\IR^{4} \times \IR^{6}/ {\Gamma}$ where the discrete
subgroup $\Gamma \subset SO(6)$
leaves the brane world-volume intact \dgmr. Then the near-horizon
geometry looks like $AdS_{5} \times S^{5}/{\Gamma}$, so that
the isometry group $SO(4,2)$ of the $AdS_5$ space still
corresponds to the conformal symmetry of the SCFT, while the
isometry of the $S^{5}/{\Gamma}$ becomes the R-symmetry group. The
field content and the interactions are nicely encoded in the
corresponding quiver diagram \refs{\dgmr,\lnv}. The gauge group is
defined by irreducible representations of $\Gamma$ \foot{The
question of whether the diagonal $U(1)$s decouple or not will
be addressed later.}:
\eqn\g{G = \prod_i \otimes SU(n_i N)}
where the product is over all the representations of dimension $n_i$.
Each arrow in the quiver diagram from the node $i$ to the node $j$
gives rise to the bifundamental matter $(n_i N, \overline{n_j  N})$.

In this paper we focus on $\CN=2$ superconformal theories
that correspond to ADE-type subgroups $\Gamma \subset SU(2)$.
Although the discussion is very general, it is convenient to think
of the particular example of the $A_{n-1} = \IZ_n$ orbifold.
The low-energy theory on the world-volume of $N$ three-branes
is then a $SU(N)^n$ gauge theory where $\IZ_n$ acts as a
permutation of the gauge factors.
The conformal dimensions of the relevant and marginal operators in
this $\CN=2$ superconformal field theory were studied in \toz.
The authors of \toz\ considered only $\IZ_n$-symmetric
operators. From the string theory point of view
these operators correspond
to the $\Gamma$-invariant, i.e. untwisted, states \dhvw.
However, because the orbifold action is not free, there are also
twisted sectors not invariant under $\IZ_n$ permutation.
Following the AdS/SCFT correspondence, these states couple to
the operators not invariant under exchange of the gauge factors.
It is convenient to choose the following basis of such operators:
\eqn\o{\CO(i) - \CO(i+1)}
where $\CO$ stands for a certain combination of chiral fields.
Some marginal deformations of the form \o\ have clear physical
interpretation corresponding to differences between coupling
constants of the $i$-th and $(i+1)$-th gauge factors \ks.

In the next section we work out the chiral primary operators
in $\CN=2$ SCFT that are not invariant under exchange
of the gauge factors and calculate their conformal dimensions.
Section 3 is devoted to the string theory analysis of twisted
states and comparison to the gauge-theoretic results. A similar
question was recently posed in the investigation of
Brane Box Models \hu. When the present paper was completed,
we received the preprint \afm\ where the arguments analogous
to our section 3.2 were applied to fields localized on
seven-branes in F-theory.


\newsec{$\CN=2$ Superconformal Gauge Theories}

We start this section by enumerating the chiral fields in the $\CN=2$
field theory with the gauge group \g\ which can be used as building
blocks for construction of the chiral primary operators. Each
node of the corresponding Dynkin diagram contributes one $\CN=2$
vector multiplet, i.e. in terms of $\CN=1,0$ chiral fields:
\eqn\fw{\Phi_i = \pmatrix{\psi_i \cr a_i} \quad
W_i = \pmatrix{A_i \cr \la_i}}
To simplify the notations we suppress
space-time indices.
In the specific case of $\IZ_n$ orbifold
there are also $n$ bifundamental matter
hypermultiplets that we write as $(Q_i, \tilde Q_i)$ in $\CN=1$
notations. The global symmetry of the $\CN=2$ gauge theory
is $SU(2)_R \times U(1)_R$.

Now using these fields we construct all the possible chiral operators
of the form \o\ whose dimensions are protected by supersymmetry.
Such operators come in short multiplets,
so that their lowest components have scaling dimensions
determined by the R-charges \shiraz:
\eqn\dimr{\Delta = \left| {R \over 2} \right| +d-1}
where $d$ is the dimension of $SU(2)_R$ representation.

It is sufficient to consider only bosonic
members of the multiplets, since the fermionic superpartners
follow by supersymmetry\foot{Actually due to the extended
supersymmetry we might consider only one representative of each
$\CN=2$ supermultiplet.}. Operators \o\ which involve
fields $Q_i$ and $\tilde Q_i$,
are not primary because they contain derivatives
of the superpotential \foot{Note the difference from
the notations of \toz\ in the relative sign.}:
\eqn\sup{\sum_i \int d^2 \th \big[ \Tr (\tilde Q_i \Phi_i Q_i)
-  \Tr (Q_{i+1} \Phi_i \tilde Q_{i+1}) \big] + c.c.}

Using the building blocks of the form $\CO = \Tr WW \Phi^k$,
$\CO = \Tr \Phi^k$ and $\CO = \Tr W \Phi^k$ we come to four
families of the "twisted" bosonic operators:

$1)$ The family of states whose lowest representative is
difference of the gluino bilinears:
\eqn\ww{\Tr (\la_i \la_i a_i^{k-1}) -
\Tr (\la_{i+1} \la_{i+1} a_{i+1}^{k-1})}
This chiral operator is a triplet with respect to the
$SU(2)_R$ symmetry and its $U(1)_R$ charge equals $2k$
where $k$ is a positive integer. Conformal dimension is
given by the classical expression $\Delta = k+2$.

$2)$ There is another bosonic state in the same supermultiplet
which corresponds to the Lagrangian density of the kinetic
term for the superfield $W$. Integration only over the part of the
superspace ensures that the states are in the short multiplet.
The chiral operator \foot{From now on tilde refers to the
Hodge dual. There will be no confusion with the fields $\tilde Q_i$
since we do not encounter the latter any more.}:
\eqn\ffm{\Tr (F_i^2 + iF_i \tilde F_i) a_i^{k-1}
- \Tr (F_{i+1}^2 + iF_{i+1} \tilde F_{i+1}) a_{i+1}^{k-1}}
couples to the difference between the $i$-th and the
$(i+1)$-th holomorphic couplings,
$\tau_j = {\theta_j \over 2 \pi} + {4 \pi i \over g_j^2}$,
while the corresponding anti-chiral operator:
\eqn\ffp{\Tr (F_i^2 - iF_i \tilde F_i) \bar a_i^{k-1}
- \Tr (F_{i+1}^2 - iF_{i+1} \tilde F_{i+1}) \bar a_{i+1}^{k-1}}
couples to the difference between the anti-holomorphic
couplings $\bar \tau_i$.
Altogether they form a complex $SU(2)_R$ singlet representation,
and their linear combinations couple to differences between
gauge couplings $g_i$ and theta-angles $\theta_i$.
The R-charges of \ffm\ and \ffp\ are equal to $\pm (2k-2)$
respectively. The conformal dimension $\Delta = k+3$ is
in agreement with the fact that \ffm\ can be
obtained from \ww\ by the action of two supercharges.

$3)$ In turn, the operator \ww\ with a fixed $k$ can be
obtained by action of two supercharges on the corresponding
difference between Coulomb moduli:
\eqn\nwa{\Tr (a_i)^{k+1} - \Tr (a_{i+1})^{k+1}}
Obviously, the state \nwa\ is a primary operator,
the lowest component of the short multiplet.
Therefore its conformal dimension $\Delta = k+1$ and
R-charges $R=2k+2$, $d=1$ obey the formula \dimr.

$4)$ The last family of chiral primary operators has the
following bosonic representatives:
\eqn\f{\Tr (F_i a_i^k) - \Tr (F_{i+1} a_{i+1}^k)}
Note that the lowest ($k=0$) state exists only if the
gauge group is $\Big(\prod_i \otimes U(n_i N) \Big) /U(1)$ and not \g.
Therefore, the presence of the corresponding states in the
spectrum of supergravity harmonics can tell us
about decoupling of the $U(1)$s.
The state \f\ is a $SU(2)_R$ singlet and carries $2k$
units of the $U(1)_R$ charge. The scaling dimension
$\Delta = k+2$ is again protected from quantum corrections.

While the operators described above manifestly comprise bosonic
content of a short $\CN=2$ multiplet in the boundary SCFT, it is
instructive to mention the supermultiplet structure of supergravity
harmonics they couple to.
Scalars {\bf 3}$_{2k}$ + {\bf 1}$_{2k-2}$ + {\bf 1}$_{2k+2}$ and
the tensor {\bf 1}$_{2k}$ naturally fall into anti-self-dual
tensor multiplet of $\CN=4$ $AdS_5$ superalgebra \gun.
Because of the multiplet shortening their masses are also
protected from quantum and stringy corrections.
Next, following arguments of \afm, it would be sufficient
to check only R-symmetry representations of Kaluza-Klein excitations
since their masses were completely determined by supersymmetry.

To conclude this section we make some predictions for
the masses of Kaluza-Klein harmonics with $R=2k$ coming from
the twisted sectors.
According to \refs{\witholo, \gkp}, the operator $\CO$
of spin zero on the boundary couples to the supergravity field
$\phi$ with the mass:
\eqn\mass{m^2 = \Delta_{\CO} (\Delta_{\CO}-4)}
For the scalar states \ww, \ffm\ and \nwa\ it means\foot{The
$k=0$ state would correspond to the non-chiral operator
$\Tr (W_i \overline{W_i}) - \Tr (W_{i+1} \overline{W_{i+1}})$ with
zero R charge.}:
\eqn\mf{m^2 = k^2-4, \quad k \ge 1}
\eqn\mb{m^2 = k^2+4k}
and
\eqn\mnwa{m^2 = k^2-4k}
correspondingly.

The eigenvalues of the Maxwell-like operator widely used
in the supergravity literature for the
tensor operator \f\ are given by \refs{\ffz, \aoy}:
\eqn\masstens{m^2 = (\Delta_{\CO}-2)^2}
which entails:
\eqn\mt{m^2 = k^2.}

For the sake of convenience we outline the expected twisted
states with a given R-charge in the following table:

\vskip 5 pt
$$
\def\tbntry#1{\vbox to 23 pt{\vfill \hbox{#1}\vfill }}
\hbox{\vrule width 1dd
      \vbox{\hrule height 1dd
            \hbox{\vrule
                  \hbox to 130 pt{
                  \hfill\tbntry{State}\hfill }
                  \vrule
                  \hbox to 100 pt{
                  \hfill\tbntry{$SU(2)_R \times U(1)_R$}\hfill }
                  \vrule
                  \hbox to 100 pt{
                  \hfill\tbntry{Mass}\hfill }
                  \vrule width 1dd
                 }
            \hrule  height 1dd
            \hbox{\vrule
                  \hbox to 130 pt{
                  \hfill\tbntry{Family 1: scalars}\hfill }
                  \vrule
                  \hbox to 100 pt{
                  \hfill\tbntry{{\bf 3}$_{2k}$, $k \ge 1$}\hfill }
                  \vrule
                  \hbox to 100 pt{
                  \hfill\tbntry{$m^2 = k^2-4$}\hfill }
                  \vrule width 1dd
                 }
            \hrule
            \hbox{\vrule
                  \hbox to 130 pt{
                  \hfill\tbntry{Family 2: scalars}\hfill }
                  \vrule
                  \hbox to 100 pt{
                  \hfill\tbntry{{\bf 1}$_{2k}$, $k \ge 0$}\hfill }
                  \vrule
                  \hbox to 100 pt{
                  \hfill\tbntry{$m^2 = k^2+4k$}\hfill }
                  \vrule width 1dd
                 }
            \hrule
            \hbox{\vrule
                  \hbox to 130 pt{
                  \hfill\tbntry{Family 3: scalars}\hfill }
                  \vrule
                  \hbox to 100 pt{
                  \hfill\tbntry{{\bf 1}$_{2k}$, $k \ge 2$}\hfill }
                  \vrule
                  \hbox to 100 pt{
                  \hfill\tbntry{$m^2 = k^2-4k$}\hfill }
                  \vrule width 1dd
                 }
            \hrule
            \hbox{\vrule
                  \hbox to 130 pt{
                  \hfill\tbntry{Family 4: 2-forms}\hfill }
                  \vrule
                  \hbox to 100 pt{
                  \hfill\tbntry{{\bf 1}$_{2k}$, $k \ge 1$}\hfill }
                  \vrule
                  \hbox to 100 pt{
                  \hfill\tbntry{$m^2=k^2$}\hfill }
                  \vrule width 1dd
                 }
            \hrule height 1dd
         }
     }
$$
\hskip 5pt

\newsec{Twisted Sectors of $\CN=2$ $AdS$ Orbifold}

The orbifold geometry is manifestly singular because of
the continuous set of fixed points -- a circle $S^1 \subset S^5$.
The twisted states are localized on this circle \dhvw,
so that the corresponding fields propagate in the
six-dimensional space-time: $AdS_5 \times S^1$.
The Kaluza-Klein harmonics of these states are not invariant
under $\IZ_n$ permutations. As explained in the introduction,
they couple to operators charged under the corresponding
discrete symmetry group.

As was shown in \dgmr, type IIB string theory on the ADE
orbifold leads to the $(2,0)$ six-dimensional effective
theory. Particularly, the twisted
sectors include as many $(2,0)$ tensor multiplets
as the number of non-trivial conjugacy classes of $\Gamma$.
The bosonic content of a tensor multiplet consists of
the anti-self-dual antisymmetric tensor $F_j$, three scalars
$\vec \xi_j$ in the triplet representation of six-dimensional
$SU(2)_R$ global symmetry and two scalar singlets
$\phi_j$ and $\vp_j$.
It is easy to see that the number of fields and their
quantum numbers indeed match the results of the previous
section if we identify $SU(2)_R$ symmetry in six and four
dimensions, and associate the four-dimensional $U(1)_R$
symmetry with rotations over the fixed circle. By this
definition, the $k$-th Fourier harmonic on the circle
carries $2k$ units of the $U(1)_R$ charge, $k \in \IZ$.

Naive Kaluza-Klein reduction on the $S^1$ gives
the masses $m^2=k^2$ for $\CN=4$ supergravity
multiplets on $AdS_5$. Except for the tensor \f, this result
differs essentially from the dimensions of the spin zero
operators \ww\ - \nwa. Below we match the Kaluza-Klein
modes from twisted sectors to these operators and find several
interesting subtleties which lead to the mass corrections.
The key difference from the flat space orbifold
$\IC^2 / \Gamma$ is due to the curvature and
the five-form flux $G^{(5)}$ through $S^5$. These background
fields induce effective interaction in the $(2,0)$
six-dimensional theory. Even though the three-point amplitude
involves only two twisted states, direct calculation
of it for the $S^5 / \Gamma$ type IIB background does
not seem promising. We choose another way and use
the blow-up of the singularity: $X \rightarrow S^5 / \Gamma$.
At least locally
we can represent $X$ as $S^1 \times \CM$, where $\CM$
is an Einstein manifold with the cosmological
constant $\Lambda =4$.
If size of $\CM$ is large enough, we can rely on
type IIB supergravity calculations.
The only non-trivial cohomology
group of $\CM$ is $H^2(\CM, \IZ)$, generated by $(n-1)$
anti-self-dual normalizable two-forms $\om_i$:
\eqn\oo{\int_{\CM} \om_i \wedge \om_j = \d_{ij}.}

With this picture in mind, let us now work out the Kaluza-Klein
spectrum for each family of fields step by step. The strategy is to
find linearized equations of motion of the $(2,0)$ theory
in six dimensions taking into account the non-trivial
background ($G^{(5)}$ and $\Lambda$).

\subsec{Family 1: $SU(2)_R$ Triplets}

The mass correction to the triplet of real scalars $\vec \xi_i$
comes from the interaction with the background curvature.
Analogous to the flat space solution, the manifold $\CM$
corresponds to $n$ Kaluza-Klein monopoles in a universe
with repulsive cosmological constant $\Lambda$. Indeed, locally
the metric on the Einstein manifold $\CM$ \gh:
\eqn\taub{ds^2 (\CM) = V^{-1} (d\tau + \vec A d \vec r)^2
+ V d \vec r^2}
resembles the metric of the Euclidean multi-centered Taub-Nut
solution. Here $V(\vec r)$ and $\vec A(\vec r )$ depend only on
the coordinate $\vec r$ on the three-dimensional base $\CB$.

The metric on $\CM$ depends on $n$ three-vectors $\vec r_i$
corresponding to the positions of Kaluza-Klein monopoles.
It is convenient to place one of the monopoles to the origin,
and choose the basis of $(n-1)$ independent positions:
$\vec \xi_i = \vec r_i$. The motion of the monopoles can
be approximated by geodesic motion on the $3(n-1)$-dimensional
moduli space \refs{\monopol,\mon}. From the six-dimensional
point of view,  $\vec \xi_i$ correspond
to the $(2,0)$ triplet of fields with the effective Lagrangian
\ruback:
\eqn\sixact{ \CL = \int_{\CM}\big[ \sqrt{g} (R-2 \Lambda)
+ G^{\a \b \gamma \d} \p_{\mu}g_{\a \b} \p^{\mu}g_{\gamma \d} \big]}
obtained via integration over $\CM$. The greek letters from
the beginning of the alphabet refer to directions along $\CM$,
and $\mu$ is one of the six-dimensional coordinates $X^{0 \ldots 5}$.
The last term in the expression \sixact\
gives the standard kinetic energy for $\vec \xi_i$, while
the former refers to the potential energy. It turns out that
the potential energy can be represented as an integral over
the base $\CB$ where it reduces to the classical expression
$\sum_i U(\vec \xi_i)$, i.e. the sum over classical
potential energies of each monopole \refs{\gh, \monopol}.
Because, to the second order in $\vec r_i$, equally charged
monopoles exert no mutual force, we end up with external gravitational
potential $U \approx 1 - \Lambda  r^2$ \ghlam.
In six dimensions
it gives the tachyonic mass $m^2=-4$ to the scalar triplet.
And the reduction on the fixed circle gives the expected
answer \mf: $m^2=k^2 -4$.

\subsec{Families 2-3: Periods of $B$ Fields}

The mass correction to scalar singlets
comes from the interaction
with the background flux $G^{(5)} = d A^{(4)}$.
These scalars are periods of the $B^{(NS)}$ and $B^{(RR)}$
two-form fields over homology two-cycles:
\eqn\decomp{B^{(NS)} = \sum_i \vp^i \wedge \om_i
\quad  B^{(RR)} = \sum_j \phi^j \wedge \om_j.}

In ten dimensions the linearized equations of motion for the $B$
fields look like:
\eqn\beq{ \nabla^{\mu} H^{(NS)}_{\mu \nu \la} =
{2 \over 3} G^{(5)}_{\nu \la \a \beta \gamma} H^{\a \beta
\gamma}_{(RR)} \quad
 \nabla^{\mu} H^{(RR)}_{\mu \nu \la} = -
{2 \over 3} G^{(5)}_{\nu \la \a \beta \gamma} H^{\a \beta
\gamma}_{(NS)}}
where $H_{(NS/RR)}$ refer to the field
strengths of $B^{(NS/RR)}$ respectively.
Because the self-dual field $G^{(5)}$ is not dynamical,
it is convenient to write an effective action for the
fields $B^{(NS)}$ and $B^{(RR)}$ that leads to the equations
\beq \dima:
\eqn\IIB{ S = \int d^{10} X \Big[
{1 \over 12} (H_{(NS)})^2 + {1 \over 12} (H_{(RR)})^2 +
4 A^{(4)} \wedge H_{(NS)} \wedge H_{(RR)} \Big]}
Using \oo\ and \decomp, we perform the dimensional
reduction of \IIB\ to six dimensions:
\eqn\six{ S= \sum_i \int_{AdS_5 \times S^1}  d^{6} X
\Big[
{1 \over 2}(d\phi_i)^2 + {1 \over 2}(d \vp_i)^2 +
4 \phi_i \wedge d \vp_i \wedge G_{(5)} \Big] }
The background flux
$G_{(5)} = \epsilon_{\mu_1 \mu_2 \mu_3 \mu_4 \mu_5}
dX^{\mu_1} dX^{\mu_2} dX^{\mu_3} dX^{\mu_4} dX^{\mu_5}$
is proportional to the volume
form on the $AdS_5$, so that the derivative in the last term of
\six\ acts in the $X^5$ direction along the $S^1$. Hence
the Fourier harmonics of $(\phi_i, \vp_i)$ propagating on
the $AdS_5$ space become mixed by the following operator:
\eqn\masmatr{\pmatrix{ \Delta(AdS_5) - k^2 & 4k \cr
4k & \Delta(AdS_5) - k^2.}}

It gives the eigenvalues of the Laplace operator $\Delta(AdS_5)$:
$m^2= k^2 \pm 4k$.
The mode corresponding to the positive sign, $m^2=k^2 + 4k$, has
exactly the same mass as expected by SCFT analysis \mb\ to couple
to the chiral operators \ffm\ of family 2.
The other harmonic has the right mass \mnwa\ to couple to
dimension $\Delta = k$ operators \nwa\ of family 3.
The negative $k$ harmonics couple to the corresponding
anti-chiral operators (e.g. \ffp) with R-charge $R=2k <0$
\foot{In this case identification of the modes is inverse:
the supergravity harmonics with $m^2 = k^2 + 4k$ couple to
anti-chiral operators of family 3 while $m^2 = k^2 - 4k$
modes couple to the states \ffp.}.

\subsec{Family 4: Antisymmetric Tensors}

The six-dimensional antisymmetric tensor $F_i$ comes from
the projection of type IIB self-dual four-form $A^{(4)}$ on the basis
of two-cycles dual to $\om_i$ \dgmr. In six dimensions the
anti-self-duality equation for the field strength $G_i=dF_i$ has
the usual form:
\eqn\teq{G_i = - \tilde G_i}

It is easy to check that to linear order this equation remains
unchanged unless we have a background flux $G$. Once the latter
does not take place, we can make further reduction on the $S^1$, and
deduce the spectrum $m^2 = k^2$ in accordance with \mt.
However, there is no massless unitary representation of
$AdS_5$ superalgebra $SU(2,2|2)$ corresponding to such state \gun.
Therefore, there is no $k=0$ mode. This means that $U(1)$ gauge
factors
decouple, and the gauge group indeed has the proposed form \g.

\newsec{Conclusions}

We derived all the relevant, marginal and irrelevant (chiral)
primary operators which couple to the twisted
states of type IIB $S^5 / \IZ_n$ orbifold.
These operators are not invariant under exchange
of the gauge factors in $\CN=2$ superconformal field theory.
The mass spectrum of
the twisted modes is obtained via reduction on the fixed circle.
Blow-up of the singularity leads to the interesting features of
monopole dynamics in presence of repulsive
cosmological constant. The Kaluza-Klein reduction gives the
correct conformal dimensions to the operators from section 2
only if the interaction with the background fields is properly
taken into account.

Because the operators constructed in section 2 do not involve
bifundamental matter, the discussion allows straightforward
generalization to non-abelian discrete subgroups $\Gamma$
along the lines of \jm.
In that case, index $i$ labels gauge factors and runs over
all the conjugacy classes of $\Gamma$. Since $\CM$ supports
as many anti-self-dual harmonic two-forms as the number of
nodes in the corresponding Dynkin diagram, the analysis of
sections 3.2 and 3.3 also remains unchanged. All the other
fields come in the same supermultiplet with the periods of
antisymmetric RR forms. Therefore, their masses also correctly
reproduce the conformal dimensions of the corresponding SCFT
operators just from supersymmetry.


\vskip 30pt
\centerline{\bf Acknowledgments}

I am grateful to M.~ Berkooz, O.~J.~ Ganor, J.~ Gomis, A.~ Mikhailov,
Y.~Oz, M.~ J.~ Strassler and  A.~M.~Uranga for helpful
conversations. Especially I would like to thank E.~Witten
for suggesting the problem and stimulating discussions.

The work was supported in part by Merit Fellowship
in Natural Sciences and Mathematics and grant RFBR No 98-02-16575
and Russian President's grant No 96-15-96939.

\listrefs

\end